# Online Verification Concept for Autonomous Vehicles – Illustrative Study for a Trajectory Planning Module

Tim Stahl[1,3], Matthis Eicher[2], Johannes Betz[1], Frank Diermeyer[1]

*Abstract*— Regulatory approval and safety guarantees for autonomous vehicles facing frequent functional updates and complex software stacks, including artificial intelligence, are a challenging topic. This paper proposes a concept and guideline for the development of an online verification module – the Supervisor – capable of handling the aforementioned challenges. The concept presented for the establishment of a Supervisor is designed in a way to identify and monitor an extensive list of features contributing to safe operation. As a result, a safe overall (sub)system is attained. Safeguarding a motion planner of an autonomous race vehicle is used to illustrate the procedure and practicability of the framework at hand. The capabilities of the proposed method are evaluated in a scenario-based test environment and on full-scale vehicle data.

## I. Introduction

Every 24 seconds, on average, someone died on the roads in 2018, according to the World Health Organization (WHO). Studies have shown that advanced driver assistance systems (ADAS) help to reduce the number of traffic accidents [1]. A further step in this direction is full automation, which aims to reduce the traffic mortality rate even more.

The research interest in the field of autonomous driving is growing and a wide range of methods has been developed. In the domain of motion planning alone, there are various search-based, optimization-based and even artificial intelligence (AI)-based approaches [2], [3]. Opposed to this development, safeguarding and approval of such complex and frequently changing software (SW) stacks becomes increasingly challenging. Relying on standard methods, approval would require a prohibitive number of test cases, and systems that are continuing their learning process at the consumer are not addressed at all. This trend indicates the need for new safeguarding and approval methods [4].

In this paper, we present a generic and structured safeguarding concept, allowing for the derivation of an online verification (OV) module for autonomous vehicles (AVs). The resulting OV module is independent of the implementation or chosen approach of the function to be monitored and can be approved by standard principles (ISO 26262 [5]). The contribution of this work is the derivation of a novel safeguarding concept able to cope

[1]Chair of Automotive Technology, Technical University of Munich, Munich, Germany
[2]Business Unit Automotive, TÜV SÜD Auto Service GmbH, Garching, Germany
[3]Corresponding author, `stahl@ftm.mw.tum.de`

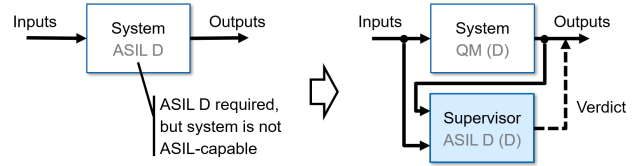

Fig. 1: Supervisor ensuring a safe overall system using the principle of ASIL decomposition (ISO 26262 [5]). A non-ASIL-capable system is monitored online for requirement violations by an ASIL D rated Supervisor.

with the following challenging properties of an AV's SW stack:

- Complex driving functions – The SW stack reaches a volume that is too large and complex for formal approval.
- Frequent updates – Every update requires a verification and validation process (at least a delta analysis).
- Machine learning methods – Learning methods might hold segments that cannot be interpreted by a human expert and are therefore hard to approve.
- Online machine learning – Systems that continuously keep learning on the fly are even more challenging since the future behavior of the system can change beyond the point of approval.

We provide a new safeguarding concept that tackles these approval issues by relying on OV. Compared to related work, the proposed concept aims for surveillance of a holistic list of properties contributing to a "safe" operation, instead of focusing on the most obvious or most safety-critical property. The Supervisor, an OV module, ensures that the (sub)system which does not meet the prerequisites for approval remains within safe boundaries (1). As a result, the overall system can be approved based on the principle of ASIL decomposition [5]. Our results are both conceptual and applied in nature. First, the concept is elaborated. Second, the method is showcased by an illustrative example of an OV module for trajectories of an autonomous race vehicle. The algorithm is evaluated based on simulative scenarios and on data-sets generated with a real race vehicle.

The remainder of this paper is structured as follows. Section II covers related work in the field. The safeguarding concept is derived in Section III and detailed in Section IV. Section V presents an illustrative example of the framework for the OV of trajectories. A discussion and concluding remarks are provided in Section VI.

## II. Related Work

With the introduction of electric devices and computing units in the automotive sector, it has become evident that those systems tend to be prone to outages or misbehavior. The ISO 26262 [5] standard offers methods to assess the potential risk and suggests appropriate safety measures for systems in motor vehicles. However, at higher levels of automation, the SW required to fulfill the driving task lacks transparency (e.g., AI [6]) and becomes too complex to meet the requirements of ISO 26262 [7], [8]. In the following, we briefly review common safeguarding approaches for the domain of AV.

One approach to guarantee safety is to examine the system under test (SUT) offline, before actual usage [9]. Model-checking or branches of modal logic are used to guarantee accordance with a specified behavior. However, frequent SW updates and online machine learning methods cannot be properly handled by these approaches, since the system may change post-examination.

Another approach is to monitor the system online to enhance the level of safety [10].

One group of approaches uses probabilistic metrics [11]–[13] to determine a collision probability or empirical performance indicators [14] resulting in a safety rating for the vehicle under test (VUT). However, a sufficient safety guarantee cannot be made based on these metrics. Furthermore, some of the trained models might extend to complex realizations, that cannot be approved themselves.

By contrast, methods relying on formal and deterministic fundamentals can provide guarantees based on imposed requirements. Among them are reachable sets [15]–[17], runtime verification [18], and metric-based approaches [19], including the Responsibility-Sensitive Safety (RSS) model [20]. However, some of the approaches are tailored to a specific SW (e.g., a certain trajectory planner) and cannot be bundled with other SW components or approaches. Furthermore, all of these approaches have in common, that they focus on selected safety aspects (e.g., dynamic collision detection) and do not strive for a holistic OV with the goal of safety approval.

## III. Safeguarding Concept

Next, the paper motivates why the concept of OV is suited to address the aforementioned challenges as well as the previously elaborated shortcomings in related work.

Koopman et al. [8] identified challenges when facing validation and verification of autonomous vehicles. The paper proposes the use of an online monitor, which fits well with the analysis in the previous section (especially when including online learning or frequent updates) and the principle of ASIL decomposition (ISO 26262). The core idea is to cross-check results generated with a non-traceable or complex algorithm (e.g., AI) using a simple and traceable monitor structure. Details about implementation specifics of such a module are not provided.

Within this paper, the terms "safe" and "safety" are interpreted in the spirit of RSS [20]. Our goal is to ensure that the VUT will not be part of the legal cause of an accident. Opposed to that, guaranteeing that an agent will never be involved in an accident is impossible.

In order to assure the safe behavior of a black-box SW component with a monitor, it must be ensured that any output generated does not result in an unsafe state. Approaches introduced in the previous section do not suffice, since only some individual features are observed. For example, if the generated trajectory of a black-box trajectory planner (e.g., AI-based [3]) is only checked for collision, other imperfections, such as exceeding the friction limit, could still result in an unsafe state.

In the next section, we incrementally deduce a guideline for the generation of an OV monitor – the Supervisor – aiming for full safety coverage of a black-box SW component, while still being approvable itself. The procedure is aligned with the core steps of the V-model, and thus also relates to the principles of ISO 26262 [5].

First, in alignment with the first step of the V-model, the requirements are specified. Here, not only are the requirements regarding the OV module specified, but the requirements for a safe functionality need to be analyzed as well. The goal is to obtain a holistic list of requirements in order to achieve full safety coverage (Section IV-A). This holistic view of safety requirements (considering the overall system) allows us to assume that once all subsystems are safe, our overall system is safe. If we were to establish these requirements internally for each module itself, this assumption would not hold.

Second, following the left side of the "V" further down, the OV architecture as well as safety measures – in line with the prerequisites of the first stage – have to be specified and implemented (Section IV-B).

Third, following the right side of the "V" back up, the developed function is integrated and tested against the requirements (Section IV-C). Finally, the established method is validated (Section IV-D).

The key steps for the generation of a holistic OV module (Supervisor – S) are summarized in Fig. 2. Four stages (S-1 to S-4) of system design, each with up to three key steps (A, B, C), have to be passed. In the following, step $y$ of stage $x$ is referred to by the notation S-$x$-$y$.

## IV. Supervisor Development Procedure

Next, the steps introduced in the previous section are elaborated in more detail.

### A. Requirements

Before the actual development of the OV module, the criteria contributing to a safe operation need to be identified. Since it is a difficult task to obtain a holistic list of criteria without unintentionally leaving out some criteria, a structured approach is proposed. This is achieved by deriving criteria from the interfaces with the surrounding environment when modeling the overall system using a systems engineering approach – as detailed in the following paragraph. However, it should be noted that this structured process supports the development towards a holistic list, but cannot provide a guarantee of completeness. Nevertheless, the final evaluation step supports keeping the probability of undiscovered vulnerabilities low.

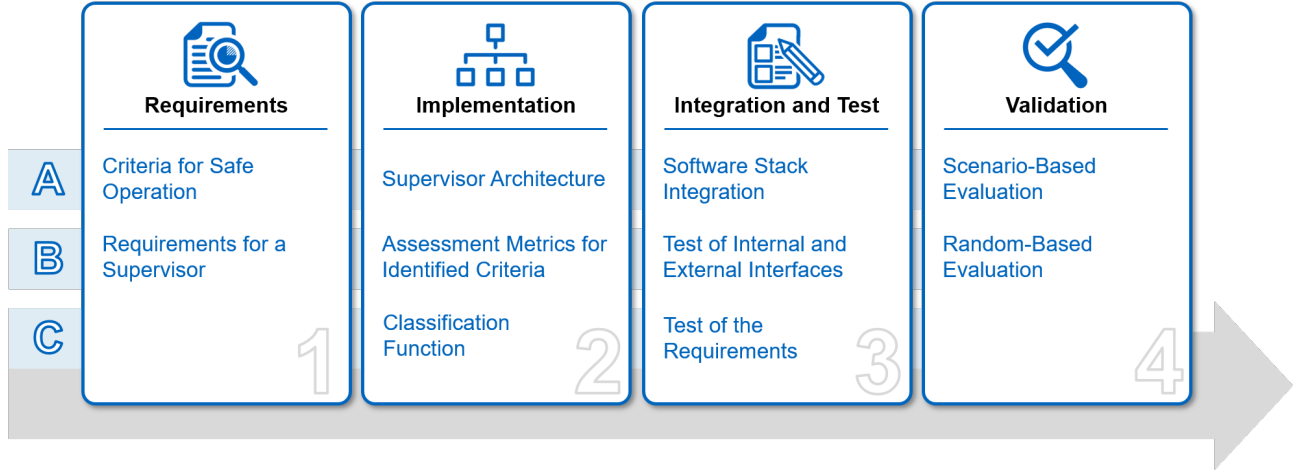

Fig. 2: Generic method for the development of an OV module. The method is structured into four stages (1-4 – indexed from left to right), each comprising up to three steps (A, B, C – indexed from top to bottom).

First, the AV and its environment is modeled by utilizing a systems engineering approach. Bagschik et al. [21] established a detailed ontology defining all environmental elements relevant for AVs. We use this information to extract the fundamental subsystems of an AV environment.[1] Details about the internal function of the system must not be known. Second, interfaces are added between any feasible combination of subsystems in the environment. For an AV, we obtained the system and interfaces depicted in Fig. 3.

Finally, the criteria for a safe operation can be derived by systematically enumerating every possible interface between the VUT and any subsystem. In doing so, each interface is analyzed for possible safety issues upon occurring imperfections. The following guiding question can be used as a basis here: "What kind of issue with this specific interface could pose a risk to safe operation?"

Besides the criteria for safe operation, the requirements of the Supervisor module itself need to be identified and tackled. Hörwick et al. [22] specified metrics to be fulfilled by OV methods. Consequently, the following requirements have to be met by a Supervisor:

- Simple and predictable behavior (to allow safety verification of the Supervisor itself)
- Real-time capability
- Separation from the overall system
- Integrity of the Supervisor (including the hardware)

### B. Implementation

After the establishment of the basic requirements, the interfaces and architecture of the verification module within the autonomous driving stack need to be identified. The end and start of the OV pipeline are tracked down based on the guidelines described below.

---
[1] All layers of the ontology are used except for L3, since L3 only holds temporary modifications of L1 and L2.

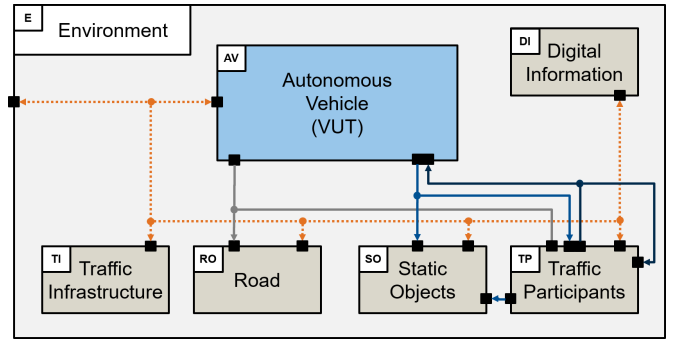

Fig. 3: System model and its interfaces based on ontology. Solid lines indicate a physical interaction and dashed lines an information flow. Directed arrows imply an interaction initiated by the subsystem at the tail.

End of the pipeline:
- Soon after the system to be safeguarded – Keep the complexity of the SW to be safeguarded as low as possible to facilitate the verification process.
- Bottleneck in the system (no parallel paths) – For full safety control, it is essential that there are no parallel ways that may shortcut the verification decisions.

Start of the pipeline:
- Output-input span – Ensure that the span of the modules to be verified is small enough, to avoid complexity and maintain the option to use established verification methods.
- Modularization – Do not start the pipeline at specific modules in order to obtain a generic OV, thereby allowing an exchange of the underlying approach or system.

Based on the defined interfaces, the architecture and information flow of the Supervisor are specified. Con-straints on the layout are set by S-1-B as well as the determined module interfaces. To fulfill the safety-related requirements S-1-B, it is necessary to cut the information flow of the original SW stack and plant the Supervisor serially. Each of the inputs interfacing the Supervisor has to align with the specification and requirements compiled from S-1-A.

For S-2-B, an evaluation metric is designed for every single criterion derived during the S-1-A requirements step. Thereby, it must follow all the other constraints obtained in the process (S-1-B and S-2-A). This step may require expert knowledge and insight into the state-of-the-art. If it is not possible to meet the requirements or provide an evaluation metric for some criteria, the architecture (S-2-A) has to be iteratively reworked.

Finally, a classification function generates a decision ("safe" or "unsafe") based on the S-2-B assessment metrics for every generated system output.

*C. Integration and Test*

After the implementation of the architecture and functions, the Supervisor is integrated into the SW stack of the AV. Verification tests of internal and external interfaces confirm a successful integration procedure. Finally, the implemented Supervisor is verified against the S-1-A and S-1-B requirements. Further guidelines and recommendations on a structured integration and verification procedure are provided in ISO 26262-4:2018,7.

*D. Validation*

To cope with design flaws affecting the safety goal, a validation test with the complete framework is required. To this end, a scenario-based and field-test validation is recommended. The scope of the scenario-based test (including fault injection principles) is to evaluate as many critical scenarios as possible, whereas field operation mainly focuses on no-fire tests. These principles form the baseline and can be extended based on the principles detailed in ISO 26262-4:2018,8.

## V. Illustrative Example: Trajectory

The method presented here is applied to facilitate OV of trajectories on an AV. More specifically, we focus on autonomous race vehicles in collaboration with Roborace [23]. Within this domain, it is possible to evaluate the function up to the dynamic limits of a vehicle and with aggressive close-proximity maneuvers.

In the following, all steps in the development stages (Section IV) are tackled individually. However, it should be noted that this simplified example only serves as an illustration of the method and does not claim completeness. A comprehensive view is part of future work.

*A. Requirements*

The system model for AVs introduced in Fig. 3 can be used for the illustrative example of safeguarding generated trajectories. All interfaces are examined with regard to possible safety-related issues related to any imperfections in the context of a trajectory. For example, the interface [AV-RO] (connecting the VUT with the road) can be an issue if the trajectory is not respecting the friction limits. When continuing this process for all interfaces, we obtain the following list of criteria (S-1-A-1 to S-1-A-6) for a safe trajectory:

1) All objects in the scene must be recognized and perceived properly ([AV-SO], [AV-TP], [AV-TI], [AV-RO], [AV-DI], [AV-E]).
2) Physical interactions with static objects or other traffic participants, initiated by the AV must not occur at any time ([AV-SO], [AV-TP]).
3) The trajectory must match the actual pose in the real world ([AV-E]).
4) The trajectory must respect the friction limits at all times ([AV-RO]).
5) Geometric and dynamic properties of the VUT must be respected at all times ([AV]).
6) Applicable rules of conduct (e.g., traffic or race rules) must be obeyed ([AV-TI], [AV-DI]).

The requirements imposed on the Supervisor do not require an application-specific analysis and can be adopted from Section IV-A.

*B. Implementation*

The system architecture step yields the following results. The end of the pipeline is selected to be any single trajectory potentially sent to the controller. The relevant modules cover all functions associated with the planning module, including the traffic prediction. Therefore, the start of the pipeline is set between the perception and the planning module (Fig. 4). The Supervisor verifies every generated trajectory against the input data. All inputs to the Supervisor must adhere to a fixed specification and the inputs originating from the perception module must be sufficiently correct at all time. The data correctness poses an assumption here and must be verified by another function or ensured by structural measures, e.g., vehicle-to-everything (V2X) communication.

It should be noted here that all relevant trajectory candidates need to be checked by the Supervisor. That way, in a framework where an emergency trajectory[2] is generated in every time step, both the driving and emergency trajectory have to be confirmed by the Supervisor. As soon as either of these two is declared unsafe, the emergency trajectory of the previous time step (declared as safe) must be executed.

As detailed in Section IV-B, every requirement identified in S-1-A has to be covered by an evaluation metric. In the following, an illustrative, albeit in some cases limited, approach for each of the identified requirements is outlined. This serves to showcase the implementation of the elaborated requirements.

---

[2]An emergency trajectory is assumed to be a trajectory ending in a safe state with zero velocity.

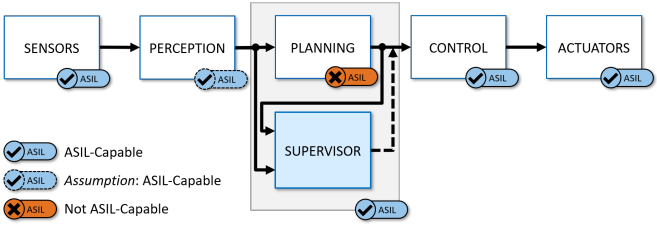

Fig. 4: Supervisor architecture for the illustrative example of a trajectory OV. The planning module is assumed to be not ASIL-capable (ISO 26262-1:2018,3.2 [5]), but with insertion of the Supervisor, the overall system can be approved.

*S-1-A-1 (Perception of objects)*: This is covered by S-2-C, since correct perception data is assumed (assured by relying on V2X data in our case).

*S-1-A-2 (Physical interactions with objects)*: This problem can be split into two subproblems – one concerned with static obstacles and one focusing on dynamic objects. A pure forward collision check along the trajectory against all static objects addresses the first subproblem.

The second subproblem is more challenging to solve. As dynamic objects decide on their future motion individually, there are infinitely many maneuver combinations. To assure absolute safety, the ego-trajectory would not be allowed to intersect (at any temporal instance) with any possible maneuvers of other vehicles. That way it would not be possible to overtake or bypass other agents. Shalev-Shwartz et al. [20] introduced a formal method relying on worst-case maneuvers (e.g., a combination of maximum braking $a_{\text{f,br}}$ of a vehicle in front, while assuming maximum acceleration $a_{\text{r,acc}}$ and worst-case braking $a_{\text{r,br}}$ for the rear vehicle). A situation is rated as unsafe when the ego-vehicle is not able to avoid contact while assuming a worst-case scenario. For instance, in the longitudinal direction, based on the inter-vehicle distance $d$, the maximal reaction time $\rho$ and the velocity of the front $v_\text{f}$ and chase vehicle $v_\text{r}$, the following constraint must be met:

$$d + \frac{v_\text{f}^2}{2a_{\text{f,br}}} - \left(v_\text{r}\rho + \frac{1}{2}a_{\text{r,acc}}\rho^2 + \frac{(v_\text{r} + \rho a_{\text{r,acc}})^2}{2a_{\text{r,br}}}\right) > 0. \quad (1)$$

For the scope of this paper, we follow the concept developed by Shalev-Shwartz. A detailed and comprehensive method will be addressed in future research.

*S-1-A-3 (Localization must be correct)*: On one hand, this condition is tackled by S-2-C, since correct perception data is assumed. On the other hand, the trajectory must host a coordinate $\langle x_i, y_i\rangle$ in a reasonable distance to the ego-pose $\langle x_{\text{ego}}, y_{\text{ego}}\rangle$[3].

*S-1-A-4 (Friction limits must be respected)*: One part of this demand is handled by S-2-C, since the determination of the actual friction limit is assumed to be provided flawlessly by the perception module[4]. The other part relies on a formal method calculating the actual combined acceleration demand along the trajectory and rates it against the maximum allowed value.

Given a trajectory (including velocity $v_{\text{x},i}$, acceleration $a_{\text{x},i}$ and curvature $\kappa_i$ at every discrete time-step $i \in N$), the total acceleration force $F_{\text{a},i}$ acting on the vehicle with mass $m$ for a time instance $i$ is given by

$$F_{\text{a},i} = m\sqrt{a_{\text{x},i}^2 + a_{\text{y},i}^2} = m\sqrt{a_{\text{x},i}^2 + (v_{\text{x},i}^2\kappa_i)^2}. \quad (2)$$

The total acceleration force should not exceed the force, the tires can counteract on the respective underground at any time. When relying on a basic tire model[5], the tires' maximal force is direction independent (ideal circle) and computed by $F_{\text{max},i} = \mu_i F_\text{N}$, where the normal force $F_\text{N}$ for a race vehicle is mainly influenced by the mass and the aerodynamic configuration. To stay on the safe side, we did not add any further aerodynamic downforce in this example. In total, the following constraint must hold:

$$\forall i \in N, F_{\text{a},i} \leq F_{\text{max},i}. \quad (3)$$

*S-1-A-5 (Geometric and dynamic properties)*: Geometric and dynamic limits of the VUT must be forward checked along the trajectory in a similar manner used for S-1-A-4. For every discrete time-step $i$ in the trajectory, the curvature $\kappa_i$ and acceleration $a_{\text{x},i}$ are compared to the vehicle's physical limits. The relevant entities are the minimal turn radius, the maximum positive acceleration (depending on engine characteristics, the drivetrain, and current velocity), and the maximum negative acceleration (depending on the braking system). It should be noted, that these constraints define the absolute bounds the vehicle is physically able to perform. In dynamic situations (especially at higher speeds), the road friction (S-1-A-4) is usually the limiting factor.

*S-1-A-6 (Rules of conduct)*: The local rules of conduct have to be formalized into mathematical constraints to be evaluated in every time step. In an urban scenario, this would be the road traffic regulations of the respective country. In the case of a race scenario, these are the race regulations in effect. In our case, only a few restrictive regulations were active that can be implemented with basic rules (e.g., maximum velocity). Future research should investigate approaches allowing rule blending and a more flexible specification using rulebooks, as introduced by Censi et al. [24].

Finally, the overall decision metric, classifying a trajectory as "safe" or "unsafe", is based on Boolean logic for the illustrative example presented here. A trajectory is rated as "unsafe" as soon as any of the constraints detailed above is violated. Consequently, an emergency trajectory rated as safe must be executed until a new trajectory that re-satisfies all constraints is generated.

---

[3] Commonly, this is one of the first coordinates within the trajectory. However, based on the path-matcher of the controller and the amount of delay and history transmitted, this may vary.

[4] This assumption might appear elusive, but one approach is to always underestimate the actual friction limit in order to guarantee feasibility.

[5] This simple model serves the concept demonstration, but can be replaced by a more advanced model.

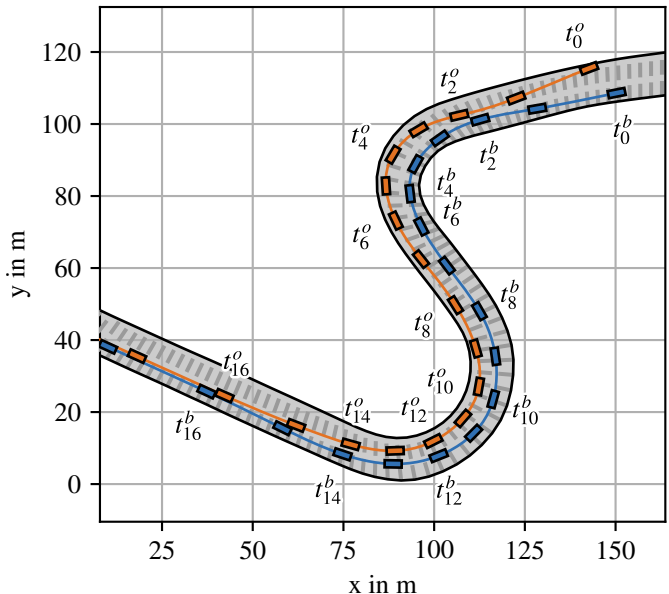

Fig. 5: Crash scenario, with the blue vehicle overtaking the VUT (orange). The vehicle poses are plotted with a temporal spacing of $1s$, where some discrete time steps within this $1s$ display are indicated by $t_i^b$ and $t_i^o$.

### C. Integration and Test

Our illustrative system was integrated into the race vehicle SW. Within the scope of this paper, we focused on fault injection and requirement-based tests. Detailed steps should be taken according to ISO 26262-4:2018,7.

### D. Validation

A validation of the overall system is carried out in order to expel possible weak spots and approve the function in favor of the safety goal. We generated a set of simulated scenarios and an interface feeding data to the Supervisor as running on the real vehicle. We generated 32 artificial critical scenarios including one or multiple vehicles [25]. Future research will investigate an extension of the scenario selection and generation process based on structured methods [26], [27]. One exemplary scenario, where another vehicle cuts off the VUT (friction and dynamics respected throughout), is envisioned in Fig. 5. When replaying the scenario, at each time step, the Supervisor is facing the blue vehicle's pose and a snippet of the ego-trajectory starting at the corresponding pose of the VUT.

As there is no unique correct system behavior for dynamic collision detection, safety envelopes in which a test passes (required latest and allowed earliest detection time) have to be defined [8]. Within our evaluation we set the safety envelopes as follows. The Supervisor must fire (rate "unsafe") no later then when a collision results while applying maximum deceleration for the VUT and assuming constant velocity for the other vehicle.

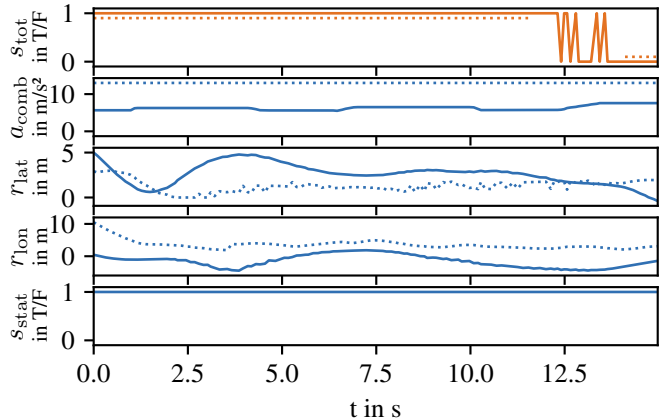

Fig. 6: Safety score $s_{\text{tot}}$ and corresponding requirement scores (combined acceleration $a_{\text{comb}}$, RSS distances $r_{\{\text{lat/lon}\}}$, static collision detection $s_{\text{stat}}$) for the presented scenario. Ground truth and bounds in dotted lines.

Furthermore, the earliest point allowed to fire is when another vehicle enters the braking distance upfront the VUT (i.e., assuming abrupt standstill). Within this interval, any rating is permitted. The preceding period must be rated as "safe" and the subsequent period as "unsafe".

The evaluation of the scenario presented here is depicted in Fig. 6. The safety score $s_{\text{tot}}$ (solid line) fits well with the defined ground truth intervals (dotted line). The scores for the static collision detection $s_{\text{stat}}$ and the combined acceleration $a_{\text{comb}}$ stayed within safe bounds the entire time. However, the RSS metrics $r_{\text{lat}}$ and $r_{\text{lon}}$ both went below the minimum required safety distance (dotted lines) at two time instances. In the first instance ($t \in [0.5s, 1.6s]$), the VUT is slightly in front of the other vehicle which is rated safe, since the race regulations state that the rear vehicle is responsible for any crash occurring. In the second instance ($t \in [12.4s, 15.0s]$), the crash risk is predicted as intended. All other evaluated scenarios (with different fault injections, e.g., exceeding friction limit or bound collision) were correctly classified according to the ground truth intervals.

Besides the scenario-based tests, we used logged field data of the trajectory planner [28] executed on a full-scale race vehicle provided by Roborace[6]. The vehicle was stable throughout the runs and safety margins to other vehicles stayed on the conservative side. The Supervisor did not fire during these runs. This observation fits well with our expectation. In conclusion, the Supervisor passed our illustrative, albeit abridged, validation procedure.

## VI. Discussion and Conclusion

In this paper, we motivated, why online verification is an appropriate method to enable safety approval for complex or non-traceable (e.g., AI-based) systems. Furthermore, we proposed a framework that guides the development process of such a holistic online verification method – the Supervisor – for an autonomous driving SW stack. The Supervisor allows approval for methods based on AI or components subjected to frequent SW updates. The concept was illustrated using the example of a trajectory OV module for an autonomous race vehicle. All steps of the development were tackled, including the creation of an extensive list of safety-relevant features for a trajectory. The scenario-based evaluation, as well as field data, provide insights with regard to validation of the overall framework.

---
[6]For this illustration we use logged data. However, a detailed validation should rely on no-fire tests directly on the vehicle.

The illustrative implementation, detailed in this paper, has limitations, especially when facing dynamic scenarios (in this example, the safety interval for the ground truth was chosen in a liberate fashion, leaving larger intervals allowing any safety rating). Ongoing research is investigating more elaborated methods for the safety assessment of collisions in a dynamic environment. Furthermore, the final version of the Supervisor module will include an extensive validation step.

ACKNOWLEDGMENTS AND CONTRIBUTIONS

Our research was supported by TÜV Süd. Tim Stahl initiated the idea of this paper and is responsible for the presented concept and implementation. Matthis Eicher, Johannes Betz and Frank Diermeyer contributed to the conception of the research project and revised the paper critically with regard to important intellectual content. Frank Diermeyer gave final approval of the version to be published and agrees to all aspects of the work. As guarantor, he accepts responsibility for the overall integrity of the paper.

REFERENCES

[1] European Commission, "Advanced driver assistance systems," Tech. Rep., 2016.
[2] B. Paden, M. Cap, S. Z. Yong, D. Yershov, and E. Frazzoli, "A Survey of Motion Planning and Control Techniques for Self-Driving Urban Vehicles," *IEEE Transactions on Intelligent Vehicles*, vol. 1, no. 1, pp. 33–55, 2016.
[3] M. Bansal, A. Krizhevsky, and A. Ogale, "ChauffeurNet: Learning to Drive by Imitating the Best and Synthesizing the Worst," in *arXiv Preprint*, 2018.
[4] H. Winner, "Introducing autonomous driving: An overview of safety challenges and market introduction strategies," *at - Automatisierungstechnik*, vol. 66, no. 2, pp. 100–106, 2018.
[5] ISO 26262, "Road vehicles - Functional safety," Tech. Rep., 2018.
[6] D. Gunning, "Explainable artificial intelligence (xai)," *Defense Advanced Research Projects Agency (DARPA), nd Web*, vol. 2, 2017.
[7] R. Salay, R. Queiroz, and K. Czarnecki, "An Analysis of ISO 26262: Using Machine Learning Safely in Automotive Software," in *arXiv Preprint*, 2017.
[8] P. Koopman and M. Wagner, "Challenges in Autonomous Vehicle Testing and Validation," *SAE International Journal of Transportation Safety*, vol. 4, no. 1, pp. 15–24, 2016.
[9] M. Luckcuck, M. Farrell, L. Dennis, C. Dixon, and M. Fisher, "Formal Specification and Verification of Autonomous Robotic Systems: A Survey," 2019.
[10] S. Lefèvre, D. Vasquez, and C. Laugier, "A survey on motion prediction and risk assessment for intelligent vehicles," *ROBOMECH Journal*, vol. 1, no. 1, p. 658, 2014.
[11] B. Kim, K. Park, and K. Yi, "Probabilistic Threat Assessment with Environment Description and Rule-based Multi-Traffic Prediction for Integrated Risk Management System," *IEEE Intelligent Transportation Systems Magazine*, vol. 9, no. 3, pp. 8–22, 2017.
[12] S. Annell, A. Gratner, and L. Svensson, "Probabilistic collision estimation system for autonomous vehicles," in *IEEE Conference on Intelligent Transportation Systems*, 2016, pp. 473–478.
[13] A. Lambert, D. Gruyer, G. S. Pierre, and A. N. Ndjeng, "Collision Probability Assessment for Speed Control," in *IEEE Conference on Intelligent Transportation Systems*, 2008, pp. 1043–1048.
[14] A. Reschka, J. R. Bohmer, T. Nothdurft, P. Hecker, B. Lichte, and M. Maurer, "A surveillance and safety system based on performance criteria and functional degradation for an autonomous vehicle," in *IEEE Conference on Intelligent Transportation Systems*, 2012, pp. 237–242.
[15] C. Pek, M. Koschi, and M. Althoff, "An Online Verification Framework for Motion Planning of Self-driving Vehicles with Safety Guarantees," *AAET*, 2019.
[16] B. Schurmann, D. Hes, J. Eilbrecht, O. Stursberg, F. Koster, and M. Althoff, "Ensuring drivability of planned motions using formal methods," in *IEEE Conference on Intelligent Transportation Systems*, 2017, pp. 1–8.
[17] M. Althoff and J. M. Dolan, "Online Verification of Automated Road Vehicles Using Reachability Analysis," *IEEE Transactions on Robotics*, vol. 30, no. 4, pp. 903–918, 2014.
[18] A. Kane, O. Chowdhury, A. Datta, and P. Koopman, "A Case Study on Runtime Monitoring of an Autonomous Research Vehicle (ARV) System," in *Runtime Verification*, ser. Lecture Notes in Computer Science, E. Bartocci and R. Majumdar, Eds. Cham: Springer International Publishing, 2015, pp. 102–117.
[19] P. Feth, D. Schneider, and R. Adler, "A Conceptual Safety Supervisor Definition and Evaluation Framework for Autonomous Systems," in *Computer Safety, Reliability, and Security*, ser. Lecture Notes in Computer Science, S. Tonetta, E. Schoitsch, and F. Bitsch, Eds. Cham: Springer International Publishing, 2017, pp. 135–148.
[20] S. Shalev-Shwartz, S. Shammah, and A. Shashua, "On a Formal Model of Safe and Scalable Self-driving Cars," in *arXiv Preprint*, 2017.
[21] G. Bagschik, T. Menzel, and M. Maurer, "Ontology based Scene Creation for the Development of Automated Vehicles," in *2018 IEEE Intelligent Vehicles Symposium (IV)*, 2018, pp. 1813–1820.
[22] M. Hörwick and K.-H. Siedersberger, "Strategy and architecture of a safety concept for fully automatic and autonomous driving assistance systems," in *IEEE Intelligent Vehicles Symposium*, 2010, pp. 955–960.
[23] J. Betz, A. Wischnewski, A. Heilmeier, F. Nobis, T. Stahl, L. Hermansdorfer, and M. Lienkamp, "A Software Architecture for an Autonomous Racecar," in *IEEE Vehicular Technology Conference (VTC2019-Spring)*, 2019, pp. 1–6.
[24] A. Censi, K. Slutsky, T. Wongpiromsarn, D. Yershov, S. Pendleton, J. Fu, and E. Frazzoli, "Liability, Ethics, and Culture-Aware Behavior Specification using Rulebooks," in *International Conference on Robotics and Automation (ICRA)*. IEEE, 2019, pp. 8536–8542.
[25] T. Stahl and J. Betz, "An Open-Source Scenario Architect for Autonomous Vehicles," in *2020 Fifteenth International Conference on Ecological Vehicles and Renewable Energies (EVER)*, 2020.
[26] A. Koenig, K. Witzlsperger, F. Leutwiler, and S. Hohmann, "Overview of HAD validation and passive HAD as a concept for validating highly automated cars," *at - Automatisierungstechnik*, vol. 66, no. 2, pp. 132–145, 2018.
[27] T. Ponn, F. Müller, and F. Diermeyer, "Systematic Analysis of the Sensor Coverage of Automated Vehicles Using Phenomenological Sensor Models," in *2019 IEEE Intelligent Vehicles Symposium (IV)*, 2019, pp. 1000–1006.
[28] T. Stahl, A. Wischnewski, J. Betz, and M. Lienkamp, "Multi-layer Graph-Based Trajectory Planning for Race Vehicles in Dynamic Scenarios," in *2019 IEEE Intelligent Transportation Systems Conference (ITSC)*, 2019, pp. 3149–3154.